\documentclass[reprint,amsmath,amssymb,aps,superscriptaddress
]{revtex4-2}

\newcounter{TRC}

\newcounter{LRLC}

\usepackage{graphicx}

\usepackage{dcolumn}
\usepackage{bm}
\usepackage{xcolor}

\begin{document}

\preprint{APS/123-QED}

\title{Molecular assembly of ground state cooled single atoms}

\author{L.~R.~Liu}
\thanks{These authors contributed equally.}
\affiliation{Department of Physics, Harvard University, Cambridge, Massachusetts, 02138, USA.}
\affiliation{Department of Chemistry and Chemical Biology, Harvard University, Cambridge, Massachusetts, 02138, USA.}
\affiliation{Harvard-MIT Center for Ultracold Atoms, Cambridge, Massachusetts, 02138, USA.}

\author{J.~D.~Hood}
\thanks{These authors contributed equally.}
\affiliation{Department of Chemistry and Chemical Biology, Harvard University, Cambridge, Massachusetts, 02138, USA.}
\affiliation{Department of Physics, Harvard University, Cambridge, Massachusetts, 02138, USA.}
\affiliation{Harvard-MIT Center for Ultracold Atoms, Cambridge, Massachusetts, 02138, USA.}

\author{Y.~Yu}
\thanks{These authors contributed equally.}
\affiliation{Department of Physics, Harvard University, Cambridge, Massachusetts, 02138, USA.}
\affiliation{Department of Chemistry and Chemical Biology, Harvard University, Cambridge, Massachusetts, 02138, USA.}
\affiliation{Harvard-MIT Center for Ultracold Atoms, Cambridge, Massachusetts, 02138, USA.}

\author{J.~T.~Zhang}
\affiliation{Department of Physics, Harvard University, Cambridge, Massachusetts, 02138, USA.}
\affiliation{Department of Chemistry and Chemical Biology, Harvard University, Cambridge, Massachusetts, 02138, USA.}
\affiliation{Harvard-MIT Center for Ultracold Atoms, Cambridge, Massachusetts, 02138, USA.}

\author{K.~Wang}
\affiliation{Department of Physics, Harvard University, Cambridge, Massachusetts, 02138, USA.}
\affiliation{Department of Chemistry and Chemical Biology, Harvard University, Cambridge, Massachusetts, 02138, USA.}
\affiliation{Harvard-MIT Center for Ultracold Atoms, Cambridge, Massachusetts, 02138, USA.}

\author{Y.-W.~Lin}
\affiliation{Department of Chemistry and Chemical Biology, Harvard University, Cambridge, Massachusetts, 02138, USA.}
\affiliation{Department of Physics, Harvard University, Cambridge, Massachusetts, 02138, USA.}
\affiliation{Harvard-MIT Center for Ultracold Atoms, Cambridge, Massachusetts, 02138, USA.}

\author{T.~Rosenband}
\affiliation{Department of Physics, Harvard University, Cambridge, Massachusetts, 02138, USA.}

\author{K.-K.~Ni}
\email[To whom correspondence should be addressed. E-mail: ]{ni@chemistry.harvard.edu}
\affiliation{Department of Chemistry and Chemical Biology, Harvard University, Cambridge, Massachusetts, 02138, USA.}
\affiliation{Department of Physics, Harvard University, Cambridge, Massachusetts, 02138, USA.}
\affiliation{Harvard-MIT Center for Ultracold Atoms, Cambridge, Massachusetts, 02138, USA.}

\date{\today}

\begin{abstract}
	We demonstrate full quantum state control of two species of single atoms using optical tweezers and assemble the atoms into a molecule. Our demonstration includes 3D ground-state cooling of a single atom (Cs) in an optical tweezer, transport by several microns with minimal heating, and merging with a single Na atom. Subsequently, both atoms occupy the simultaneous motional ground state with 61(4)\% probability. This realizes a sample of exactly two co-trapped atoms near the phase-space-density limit of one, and allows for efficient stimulated-Raman transfer of a pair of atoms into  a  molecular bound state of the triplet electronic ground potential $a^3\Sigma^+$.
	The results are key steps toward coherent  creation of single ultracold molecules, for future exploration of quantum simulation and  quantum information processing.
\end{abstract}

\maketitle

\section{Introduction}

Building up complex many-body systems from simpler, well-understood constituents is a promising approach toward understanding and controlling quantum mechanical behavior. Using ultracold molecules as  building blocks would allow new explorations of  quantum chemical dynamics~\cite{Krems2008}, 
novel quantum many-body phases~\cite{Baranov2012}, and quantum computation~\cite{DeMille2002, Andre2006, Ni2018}.

These prospects hinge on the precise generation and control of ultracold molecules with well defined internal and motional quantum states.
Many approaches for trapping and cooling molecules to ultracold temperatures are being pursued~\cite{carr2009,Ni2008, Danzl2008, Lang2008, Chotia2012, Takekoshi2014, Molony2014, Park2015, Guo2016, Norrgard2016, Prehn2016}. Recent highlights  include  rapid progress  made with laser cooling  of molecules~\cite{Barry2014, Truppe2017,Cheuk2018,Collopy2018}, the creation of quantum degenerate gases of fermionic KRb~\cite{Demarco2019}, assembling single molecules~\cite{Liu2018} and loading single molecules into an optical tweezer array~\cite{Anderegg2019}.  Bulk samples of ultracold molecules have already proven a versatile platform, enabling the study of ultracold chemistry~\cite{Ospelkaus2010} and quantum spin models~\cite{Yan2013}. In addition, strongly interacting phenomena can be explored with a lower entropy gas and with single molecule addressability~\cite{Gadway2016}. Inspired by techniques in optical single atom manipulation~\cite{Schlosser2001, Miroshnychenko2006, Kaufman2012,Thompson2013,Xu2015,Yu2018}, we proposed a realization 
through single particle control of molecules without relying on collisions for cooling~\cite{Liu2017}. 

In this paper, we experimentally demonstrate  key steps toward such an ``ultracold molecular assembler"~\cite{Liu2017}. We obtain full quantum state control including cooling, transport, and merging of two different single atoms. We perform two-photon dark resonance spectroscopy to locate the least-bound NaCs molecular state of the electronic triplet ground potential  $a^3\Sigma^+$.

  We then transfer two single ground state-cooled atoms in the same tweezer to the weakly-bound molecular state using a two-photon Raman pulse.  In the following sections, we detail each experimental assembly step.

\begin{figure}
	\includegraphics[width=\columnwidth]{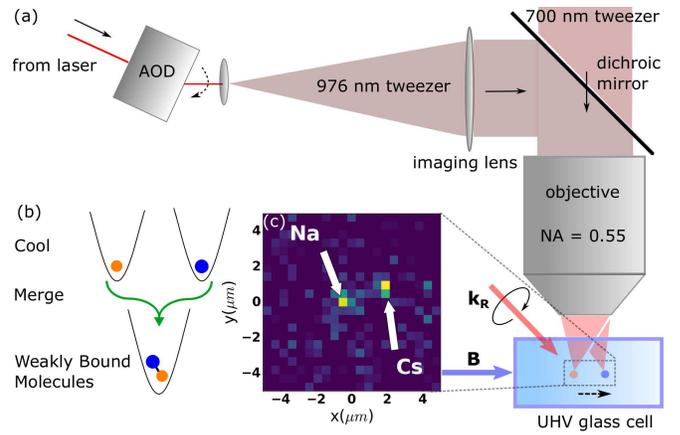}
	\caption{\textbf{Single atom trapping and  transport for molecular assembly.} \textbf{(a) Schematic of apparatus.} Two neighboring optical tweezers at 976~nm and 700~nm trap a single Cs (blue) and Na (orange) atom in the vacuum chamber. Both tweezer beams are combined on a dichroic mirror and focused by an objective. The 976~nm tweezer can be moved in the focal plane by changing the drive frequency of an upstream acousto-optic deflector (AOD).  Once atoms are cooled and merged into the same tweezer, a laser propagating along $\mathbf{k_R}$ transfers them into a bound molecular state in the presence of a  quantization \textbf{B} field. \textbf{(b) Experimental assembly steps of ultracold molecules demonstrated in this paper.}  A single Na and Cs atom are cooled, merged into the same trap, and transferred to a weakly-bound molecule.  \textbf{(c) Single-shot fluorescence image of single Na and Cs atoms in adjacent tweezers separated by 3 $\mu$m.}} 
	\label{app}
\end{figure} 

\section{Controlling the Quantized Motion of Atoms}
\label{sec:cool}

A schematic of the apparatus is shown in Fig.~\ref{app}. As described in our previously work~\cite{Liu2018}, we generate two tweezer traps at different wavelengths for quasi-independent manipulation of single Na and Cs atoms. One of the beams is steerable, so that initially separate tweezer traps can be merged. Single-atom fluorescence images confirm simultaneous trapping of single Na and Cs atoms side-by-side as shown in Fig.~\ref{app}.

Using standard polarization gradient cooling (PGC), it is possible to cool the motion of single Cs or Na atoms to an average of tens of quanta in a tight tweezer trap.  
To further cool the atoms into the lowest motional state, we use 3-dimensional Raman Sideband Cooling (3D RSC), first demonstrated with single ions~\cite{Monroe1995} and more recently with single neutral atoms~\cite{Li2012, Kaufman2012, Thompson2013}. We operate in the resolved sideband regime where the linewidth of the cooling transition is less than the trap frequency (10-100's of kHz). 
We have previously demonstrated ground-state cooling of single Na~\cite{Yu2018}. Here, we demonstrate 3D RSC of a single Cs atom in an optical tweezer. To our knowledge, we report the highest 3D ground-state probability for single atoms in tweezers to date. 

The RSC sequence consists of two steps: a coherent two-photon Raman transition that connects two internal states while removing a motional quantum, and an optical pumping (OP) step that re-initializes the internal state of the atom. The two steps are repeated until the atom reaches the motional ground state.

\begin{figure*}
	\includegraphics[width=\textwidth]{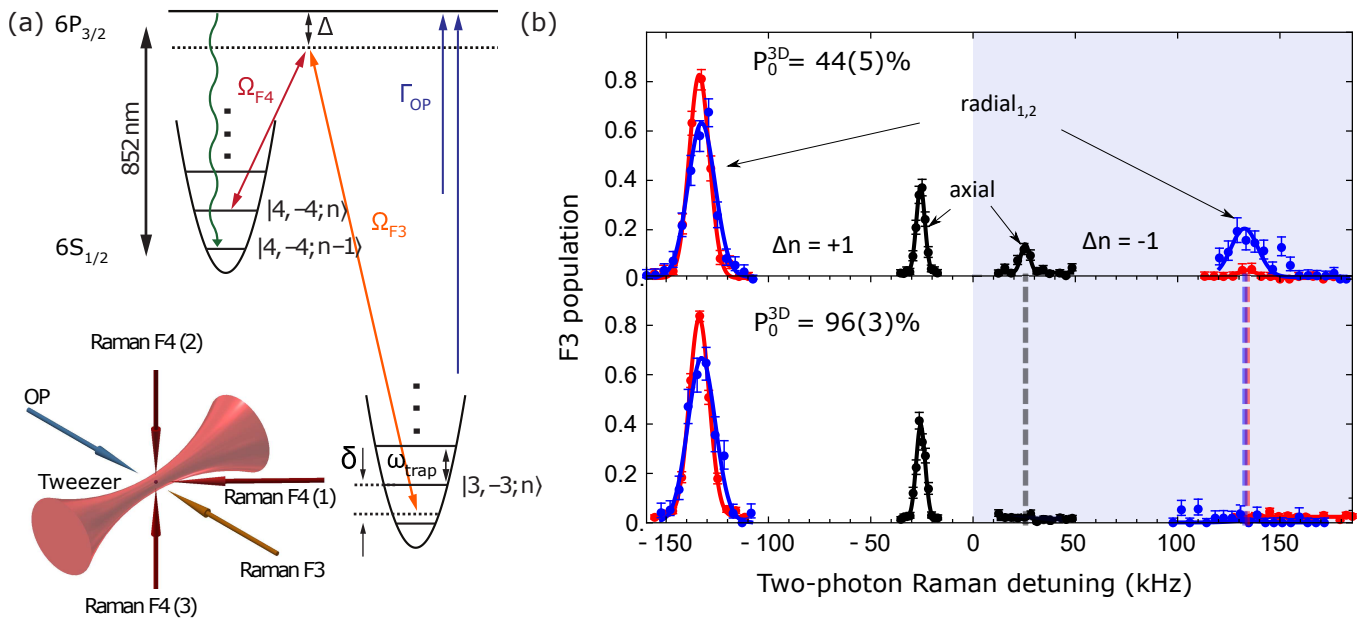}
	\caption{\textbf{3D motional control of a single Cs atom.} \textbf{(a) Level scheme for Cs RSC.} F3 and F4 Raman beams coherently couple adjacent motional states to reduce motional energy, while optical pumping provides the dissipation needed for cooling. \textbf{[Inset] Directions of laser beams.} Switching Raman F4 beam directions allows addressing motion in all 3 dimensions. \textbf{(b) 3D sideband thermometry for Cs after RSC.} Black, blue, and red spectral peaks in the unshaded (shaded) region correspond to $\Delta n = +1(-1)$ sidebands for the axial and two radial directions, respectively. Above: Spectra after sub-optimal RSC reveals the $\Delta n = -1$ sidebands, and hence the motional frequencies. The 3D ground state population is $P_0^{3D}=44(5)\%$. Below: Spectra after cooling with optimized motional frequencies, yielding $P_0^{3D}\geq96(3)\%$.}
	\label{rsc_scheme}
\end{figure*}

In our scheme (Fig.~\ref{rsc_scheme}),  the Raman transition occurs between Cs ground-state hyperfine levels   $|F=4,m_F=-4;n\rangle$ and $|3,-3;n-1\rangle$, which are about 9.2 GHz apart. Here, $n$ is the motional quantum number. The transition is driven by two phase-locked diode lasers, ``F3'' and ``F4'', both red-detuned by $\Delta=2\pi\times44$ GHz from the Cs $D_2$ line at 852~nm, and with Rabi rates $\Omega_{F3}$ and $\Omega_{F4}$, respectively. The tweezer has a power of 14.3~mW and beam waist of 0.84~$\mu$m.
To achieve motional coupling, the laser beams are arranged as shown in the inset of Fig.~\ref{rsc_scheme}(a). This configuration yields substantial two-photon momentum transfer, $\Delta \vec{k} = \vec{k}_{F4(i)}-\vec{k}_{F3}$, while the energy difference associated with the hyperfine level and motional state change is supplied by their relative detuning, $\delta$. This resonance condition is maintained for all relevant motional states, $n$. 

The atom is initially prepared in $|4,-4\rangle$ by OP (independently of the motional state, $n$). For this, we use $\sigma^{-}$-polarized beams resonant with $|4,-3\rangle \rightarrow |4',-4\rangle$ and $|3,-3\rangle \rightarrow |4',-4\rangle$ transitions, where the primed levels denote sub levels of the $6P_{3/2}$ manifold of Cs. During the first step of RSC, a Raman $\pi$-pulse drives the transition $|4,-4;n\rangle\rightarrow |3,-3; n-1\rangle$.  Subsequently, OP pumps the atom to $|4,-4;n-1\rangle$. 
OP preserves the motional state with high probability. Thus, in each RSC cycle, $n$ decreases on average. The process repeats until the atom reaches the dark state $|4,-4;0\rangle$, thereby deterministically preparing the internal and the motional quantum state of the atom. 

We switch between the three Raman F4($i$) directions in the sequence ${i=3,1,2,1}$ to cool the atomic motion along all three axes of the tweezer. The tweezer potential has a cigar shape with two near-degenerate, tightly confined. radial directions and a loosely confined axial (along the tweezer beam propagation) direction. 

The linewidth of the Raman transition is Fourier broadened due to the finite duration of a $\pi$-pulse, which is inversely related to the peak effective Raman Rabi rate  $\Omega_{F3}\Omega_{F4}/2\Delta =2\pi\times33~$kHz ($2\pi\times7\,$kHz) for radial (axial) trap axes. The smaller energy splitting of the axial motion necessitates a smaller Raman coupling along that direction. 
An 8.6~G magnetic field is applied throughout RSC along the OP propagation direction to define the quantization axis.

All Raman pulses in this experiment for cooling and spectroscopy use a Blackman window temporal intensity profile to reduce off-resonant excitation of the carrier. The starting temperature of $9.2~\mu$K, corresponding to a mean axial motional quantum number $\bar{n}_a=9$, leads to non-negligible occupation of levels up to $n_a\approx 40$. 

Due to the $\sqrt{n}$ scaling of sideband transition strengths~\cite{wineland1998}, it was necessary to ``sweep" the Raman pulse durations in descending order starting from $n_a^{init}$=41. Furthermore, to overcome decoherence, which reduces the transfer fidelity of each pulse, we repeat the sweep, but each time with a smaller $n_a^{init}=\left\{41,31,16,11,6\right\}$. The entire process takes $\approx100$~ms.

We characterize two cooling experiments in Fig.~\ref{rsc_scheme}(b): (1) sub-optimal cooling was used with slightly off-resonant $\delta \neq \omega_{trap}$ to reveal the location of the $\Delta n = -1$ sidebands. (2) optimal cooling is obtained by setting $\delta = \omega_{trap}$, as determined by the sideband locations in (1).

To characterize the cooling performance, we use sideband thermometry~\cite{Monroe1995}. Following RSC, we measure the ratio of $\Delta n=-1$ and $\Delta n=+1$ Raman sideband transition heights. A successful transition changes the state from $|4,-4\rangle$ to $|3,-3\rangle$ and is revealed by state selective imaging: light that is resonant with the cycling $|4,-4\rangle \rightarrow |5',-5\rangle$ transition ejects only $|4,-4\rangle$ atoms. The remaining atoms in $|3,-3\rangle$ are then imaged.
We obtain the average occupation number $\bar{n}$ from the ratio of sideband heights via $I_{-1}/I_{+1} = \frac{\bar{n}}{\bar{n}+1}$. By assuming a thermal distribution, we extract a temperature and a ground state probability along each axis. The product of the ground state probabilities in all three dimensions gives the 3D ground state probability $P_0^{3D}$.

This procedure yields $\{\bar{n}_a, \bar{n}_{r1},\bar{n}_{r2}\} = \{0.03(3),0.00(1),0.01(1)\}$, corresponding to $P_0^{3D}\geq96(3)\%$ for optimal cooling.

The signal contrast in Fig~\ref{rsc_scheme}B does not reach unity due to the $\approx300~\mu$s coherence time for driving motional sideband transitions. Furthermore, different pulse durations were used on the two radial axes, leading to a further difference in contrast. However, the sideband \textit{ratios}, used to extract the final ground state population, are unaffected.  

A final consideration is that any wait time between the end of RSC and molecule formation needs to be minimized because the atoms can be heated by off-resonantly scattering photons from their respective tweezers. This occurs at a rate of $\dot{\Delta n_a}\approx0.3$ Hz. To avoid unnecessary waiting, we perform the Na and Cs RSC sequences concurrently  so that they end at the same time. We have verified experimentally that RSC of one species does not affect the atom of the other species.

\section{Preparing both Na and Cs in the ground state of the same tweezer}
\label{joint_heat_section}

As shown in Fig.~\ref{app}, two optical tweezers trap  a single Cs  and Na atom approximately 3 $\mu$m apart. Both tweezer beams are combined on a dichroic mirror and focused by a NA$=0.55$ objective. The position of the Cs tweezer can be moved by changing the drive frequency of an upstream acousto-optic deflector (AOD).   While merging two separately confined identical ground-state atoms into one potential well requires delicate quantum tunneling~\cite{Kaufman2014}, merging different atomic species is more straightforward.  Due to their different atomic polarizability  as a function of wavelength, two different color optical tweezers allow the two atoms to be manipulated quasi-independently~\cite{Liu2018}.  

\begin{figure*}
	\includegraphics[width=\textwidth]{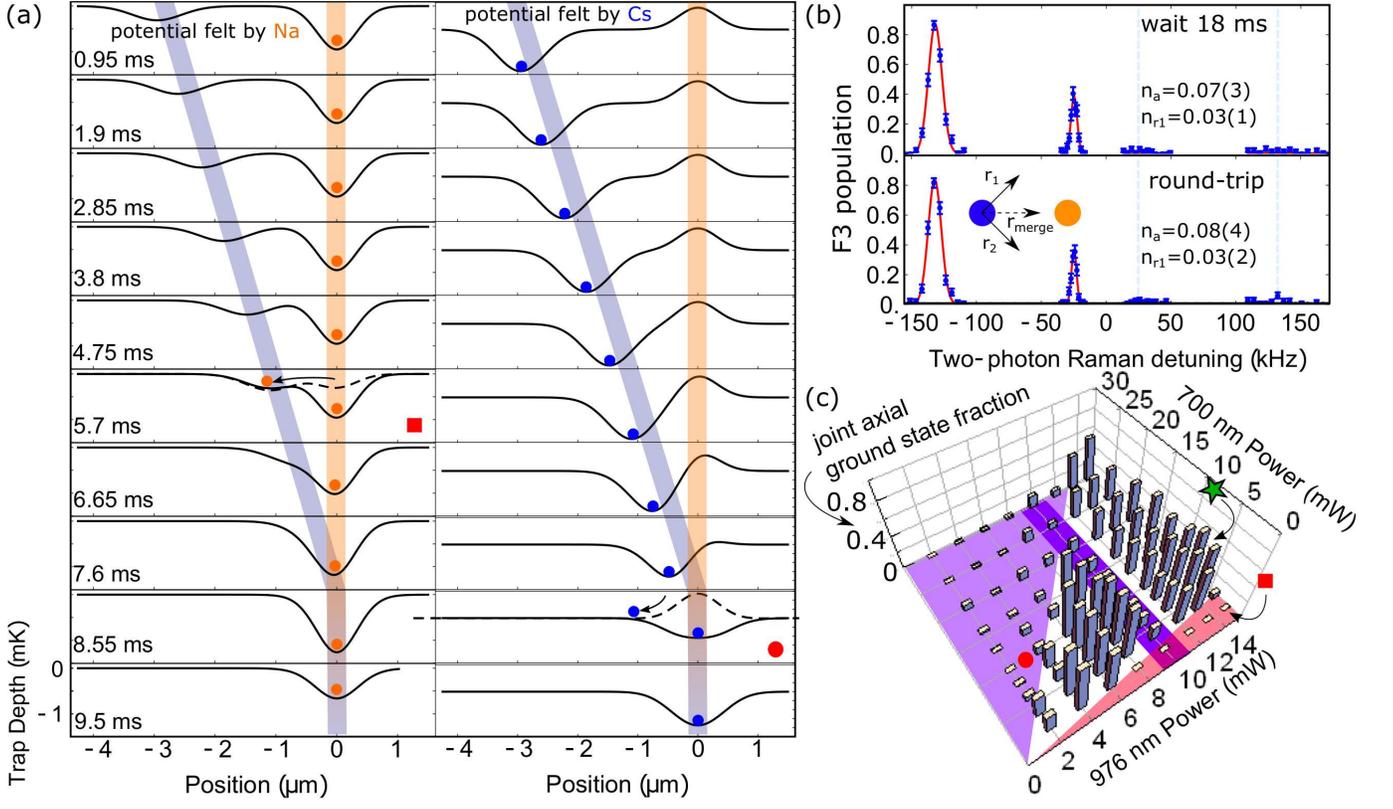}
	\caption{\textbf{Merging atoms in two tweezers while maintaining  quantum motional states.} \textbf{(a) Radial cuts of optical potential experienced by Na and Cs during the merge time sequence.} Blue and orange lines show paths of the 976~nm and 700~nm tweezers, respectively. The 976~nm tweezer containing Cs is translated by $2.95~\mu$m in 7.6~ms until it overlaps with the 700~nm tweezer. Then, the 700~nm tweezer power is linearly ramped from 48~mW to 0~mW in 1.5~ms, followed by a 50~$\mu$s wait. Dashed potential in the left 5.7~ms panel (marked by red square) shows the conditions for non-ideal tweezer powers, leading to spilling of Na. Dashed potential in the 8.55~ms panel (marked by red circle) shows the conditions for a different set of non-ideal tweezer powers giving rise to anti-trapping for Cs. \textbf{(b) Raman sideband spectroscopy to characterize heating associated with atom transport.} Top: A control experiment holding the atoms stationary for 18~ms. Bottom: After the \textit{round-trip}  merge sequence (the sequence shown in (a) followed by its time reverse) Dashed blue lines indicate expected position of $\Delta n = -1$ sidebands. The \textit{round-trip}  sequence causes minimal heating.  Inset: Coordinates of the transport direction vs a thermometry axis. Blue and orange circle represent 976~nm and 700~nm tweezers respectively. \textbf{(c) Na+Cs joint axial ground state fraction after \textit{round-trip} merge sequence as a function of 700~nm and 976~nm tweezer powers.} The lower triangle in red corresponds to spilling of Na. Red square is an exemplary point in this regime, whose radial potential is plotted with a dashed line in the correspondingly marked panel in (a). Upper triangle in purple indicates anti-trapping of Cs. Red circle is an exemplary point, whose  potential is plotted with a dashed line in the correspondingly marked panel in (a). Dark purple stripe shows parametric heating resonance (due to technical imperfection) during transport of Cs. Our usual operating point is indicated by the star.}
	\label{merge}
\end{figure*}

One tweezer beam at a wavelength of 700~nm confines Na at the intensity maximum while repelling Cs. A second tweezer beam at a wavelength of 976~nm strongly confines Cs while weakly attracting Na.  As shown in Fig.~\ref{merge}(a), translation of the 976~nm beam to overlap the 700~nm beam, followed by gradual turn-off of the 700~nm beam leaves the two atoms confined in the same tweezer trap, all within 10~ms. The exact trajectory is detailed in Appendix~\ref{app:merge_trajectory}.

After running this sequence (in the absence of Na) followed by its time reverse for detection, Raman sideband thermometry on the separated tweezer shows minimal motional excitation of Cs ($\{\Delta \bar{n}_a,\Delta \bar{n}_{r1}\}  = \{0.01(5), 0.00(2)\}$) (Fig.~\ref{merge}(b)).  

We further explore different trap powers for merging of Cs and Na atoms into one tweezer.
To prevent spin-changing collisions~\cite{Liu2018}, we first prepare Na in $|2,2\rangle$ and Cs in $|4,4\rangle$. Then, we merge the atoms and measure the joint axial ground state fraction $P_{n_a=0}^{Na}\times P_{n_a=0}^{Cs}$ as a function of beam powers (Fig.~\ref{merge}(c)). We identify three issues that can cause excess heating during the merge and require careful beam-power selection to overcome:

\begin{enumerate}
	\item The 976 nm beam can make Na spill from the 700 nm tweezer and gain kinetic energy.  This limits the ratio $P_{\mathrm{700nm}}/P_{\mathrm{976nm}}$ to be above 0.37, indicated by the  red triangle in Fig.~\ref{merge}(c), and the left panel at 5.7~ms in Fig.~\ref{merge}(a).
	\item The 700 nm beam can dominate the 976 nm beam and repel Cs from the trap.  This limits the power ratio of the beams $P_{\mathrm{700nm}}/P_{\mathrm{976nm}}$ to be below 2.7, indicated by the left purple shaded triangle in Fig.~\ref{merge}(c), and the right panel at 8.55~ms in Fig.~\ref{merge}a.
	\item   Modulation in the tweezer power during trap movement causes parametric heating of the atoms.   A weak acoustic standing wave in the AOD crystal results in an overall efficiency that modulates $\sim 1.5\%$  with the acoustic drive frequency.
\end{enumerate}

We choose powers of $P_\mathrm{976nm}=14.3$~mW  and $P_{\mathrm{700nm}}=$7.1~mW (also used in Fig.~\ref{merge}(b)) for all subsequent experiments. These powers yield the trap potentials depicted by the solid lines in Fig.~\ref{merge}(a)), i.e., approximately 2~mK for Cs and 1~mK for Na, respectively. We characterize with 3D Raman sideband thermometry that we have prepared two atoms in the same tweezer with a phase space density (PSD) of $P_0^{Na}\times P_0^{Cs}=0.80(3)\times 0.76(4)= 0.61(4)$.   In this experiment, lower optical pumping fidelity resulted in a higher initial Cs temperature as compared to Sec.~\ref{rsc_scheme}.

\section{Two-photon Raman transfer to the least-bound molecular ground state }

\begin{figure*}
\centering
	\includegraphics[width=\textwidth]{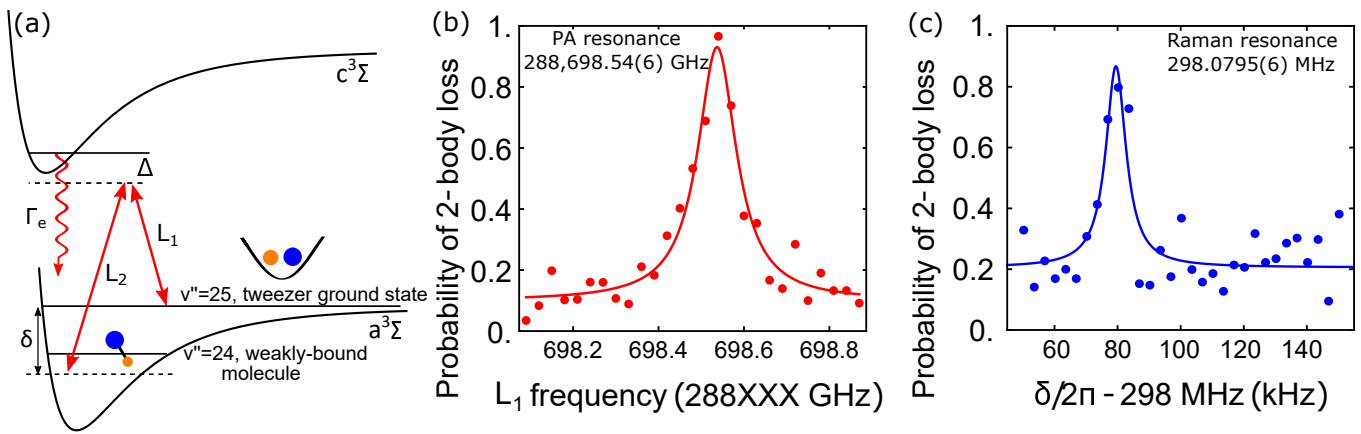}
	\caption{\textbf{(a) Level diagram for two-photon Raman transfer from an atom pair to a weakly-bound molecule.} Two lasers $L_1$ and $L_2$ with a frequency difference $\delta$ derived from an AOM are phase coherent and drive the atoms from the tweezer motional ground state  ($v''=25$) to the weakly-bound molecular state $a^3\Sigma^+(v''=24$ or $v''=-1)$.  A large detuning $\Delta$ from  the excited state $c^3\Sigma(v'=0)$, which decays at a rate $\Gamma_e$, 	reduces spontaneous emission during molecular transfer, which occurs when the two-photon frequency difference $\delta$ is resonant with the binding energy.
    \textbf{(b) Photoassociation spectroscopy of the intermediate excited state.}  With the laser $L_2$ off, the $L_1$ laser drives the atoms to the excited molecular vibrational states when resonant.    The  Na + Cs two-body loss probability is measured as a function of the PA frequency for a 75~ms pulse duration.  The $c^3\Sigma^+_1(v'=0,J'=2,F'=7)$ state is observed at 288,698.64(6) GHz.
	\textbf{(c) Two-photon Raman resonance for transferring single atoms to a molecule.} With a detuning $\Delta=2\pi\times 3.2$~GHz,  the frequency difference $\delta$ of the $L_1$ and $L_2$ beams  is scanned around the binding energy of the $a^3\Sigma^+(v''=24)$ state.    The Raman resonance is observed  at 298.0795(6)~MHz with a  FWHM of 8(2)~kHz, indicating transfer of the atom pair to the weakly-bound molecular state. }
	\label{raman_transfer}
\end{figure*}

After the merge,  both Na $|F=2,m_F=2\rangle$ and Cs $|4,4\rangle$ atoms occupy the motional ground state of the same tweezer, which corresponds to the vibrational level $v''=25$ of the combined tweezer potential and  electronic ground  molecular potential $a^3\Sigma^+$~\cite{Liu2017}.  
As shown in Fig.~\ref{raman_transfer}(a), we transfer the atom pair into the least-bound molecular state  $v''\!=\!-1$ (or $v''=24$) via a  two-photon Raman pulse using two beams $L_1$ and $L_2$ that couple the initial and final states to a single intermediate electronic excited state.

 The creation  of a molecule in the weakly-bound state is an important step toward subsequent transfer to deeply-bound molecular states via Stimulated Raman Adiabatic Passage (STIRAP)~\cite{Danzl2008, Ni2008, Lang2008, Chotia2012, Takekoshi2014, Molony2014, Park2015, Guo2016, Norrgard2016, Prehn2016}. In previous work, weakly-bound molecules were produced using Fano-Feshbach resonances.  In this work, we instead use an all-optical technique~\cite{Rom2004} to generalize the weakly-bound molecule production to atoms without suitable Fano-Feshbach resonances.

We choose $v'\!=\!0$ of the molecular potential $c^3\Sigma^+$ as the excited intermediate  state 
because it has suitable Franck-Condon factors (FCF's) with both the initial and final states, and because its large detuning from the near-threshold and trap states minimizes their contribution to spontaneous emission~\cite{Liu2017}. We choose the weakly-bound ground state $v''\!=\!-1$ as our target state because its FCF is the most similar to that of the motional ground state ($v''\!=\!25$). The spontaneous emission from the excited state during Raman transfer is proportional to the ratio of the FCF's of the two ground states to the excited state.  While a STIRAP pulse sequence could also potentially be used for this transfer, we have found from simulations that large Stark shifts of the two-photon detuning and the longer required duration result in poor efficiency. 

Initial search for the intermediate excited state relied on photoassociation (PA) spectroscopy of the two atoms. Guided by the $c^3\Sigma^+$ potential curve from Ref.~\cite{Grochola2011},  we scanned the frequency of a tunable diode laser around 1038 nm wavelength ($L_1$) until the laser was resonant with the excited state, and molecule formation was indicated by simultaneous atom loss. 

Specifically, after illuminating the atoms for 75~ms with 15 mW of $\sigma^+$ polarized light and a beam radius of $w \approx 15 $ $\mu$m, the atom merge sequence was immediately reversed to separate the surviving atoms for detection.  The two-body loss spectrum is shown in  Fig.~\ref{raman_transfer}(b).  The Lorentzian fit gives a transition frequency of 288,698.54(6)~GHz, which we identify as the $c^3\Sigma_{\Omega=1}^+ (v'=0,J'=2,F'=7)$ state, where $J'$ is total angular momentum excluding nuclear spin,  and $F'$ is the total angular momentum including nuclear spin.  The uncertainty is dominated by the wavemeter inaccuracy of 60 MHz.   We also observe the  
$J'=1$ and $J'=3$ rotational lines and fit them to $v_{J'}=v_0+ B J'(J'+1)$ to obtain a rotational constant of $B= 1.1$~GHz  The lack of a $J'=0$ state confirms  $\Omega=1$.    

To achieve two-photon Raman transfer to the ground molecular state, we first located the least-bound state $a^3\Sigma^+ (v''=24, N''=0, F''=6)$ via dark-resonance spectroscopy and calibrated the single-photon Rabi rates of the two individual beams (Appendix~\ref{app:calibrate_omega1_omega2}).  We then  increased the detuning $\Delta$ in order to reduce population of the excited state, which decays rapidly. 

Figure~\ref{raman_transfer}c shows the Raman resonance for a pulse length of 100~ms and $\Delta=2\pi\times3.2~$GHz. The two beams $L_1$ and $L_2$ propagate along $\mathbf{k_R}$ as shown in Fig.~\ref{app}, with beam radii  $\{w_0^x,w_0^y\}=\{10,23\}~\mu$m and identical beam powers of 15~mW to minimize scattering (Appendix~\ref{app:equal_beam_powers}).  The resonance is fit to a Lorentzian centered at 298.0795(6)~MHz with a FWHM of 8(2)~kHz. The 70(10)\% transfer efficiency matches closely to the relative ground-state fraction of the Na+Cs atom pair, while the $21(2)\%$ background level can be explained by  spontaneous Raman scattering of the tweezer light from the $v''=25$ state, followed by a spin-changing collision~\cite{Liu2018}. 

We have not yet observed coherent atom-molecule oscillations between the initial and final state and believe the main source of decoherence is off-resonant scattering of the Raman light from the least-bound molecular state. For the above conditions, this scattering rate is $\Gamma_{\text{Raman}}\approx 149~$Hz, larger than the Raman transfer rate $\Omega_R=2\pi\times50~$Hz. Although increasing the detuning $\Delta$ improves the ratio of Raman transfer to scattering rate, the fixed scattering rate of  $\Gamma_{\text{tweezer}}=30~$Hz due to the tweezer (Appendix~\ref{app:tweezer_v24_scattering_rate}) provided a further constraint. 

A potential solution for future work is replacing the $976~$nm tweezer with a 1038~nm tweezer that can also serve as the molecular transfer beam. Due to the tight focusing of the tweezer, the product $\Omega_1 \Omega_2$ can be more than 200 times higher for the same beam power, thereby allowing $\Delta$ to increase to reduce off-resonant scattering, while maintaining the same $\Omega_R$. 

\section{Summary and outlook}
We have described experimental steps towards coherent assembly of single molecules from individual atoms.  Starting with side-by-side trapping of the constituent atoms (Cs and Na) in optical tweezers, we have demonstrated  ground-state  cooling of Cs to its 3D ground state (96~\%) and merging single Cs and Na atoms into the same tweezer while maintaining both atoms in the motional ground state with 61~\% probability. 
These tools of dual-species single-atom manipulation can be extended to other species and tweezer wavelengths, providing a 
valuable resource to investigate interactions, collisions, and  coherent spectroscopy and creation of molecules.

With two atoms in a tweezer, we have probed their electronic ground and excited molecular potentials. The resulting information enabled two-photon Raman transfer of 70~\% of the atom pairs into the least bound molecular state of the triplet ground electronic  potential $a^3\Sigma^+$. 
In the future, deriving the molecular transfer and tweezer beams from a single laser may reduce off-resonant scattering of the transferred molecule, which is otherwise long-lived.  The transfer from the weakly-bound state to the ro-vbirational ground state could then be achieved by performing STIRAP with an excited state from the mixed potentials $B^1\Pi$ and $c^3\Sigma$.

For studies of ultracold chemistry, quantum information, and many-body physics, the number of atom pairs could be scaled up by employing an array of single atom tweezer traps as a starting point~\cite{Barredo2016, Endres2016}.

\begin{acknowledgments}
	This work is supported by the Arnold and Mabel Beckman Foundation, as well as the NSF (PHYS-1806595), the AFOSR Young Investigator Program, the Camille and Henry Dreyfus Foundation, and the ARO DURIP (W911NF1810194). J.T.Z acknowledges support from the NDSEG fellowship. 
\end{acknowledgments}

\appendix

\section{Trajectory for merging two atoms into one tweezer}
\label{app:merge_trajectory}
The speed at which we choose to transport a Cs atom in the 976~nm tweezer and subsequently merge it with the 700~nm tweezer is constrained by two main factors: (1) minimizing heating due to jerk (time-derivative of acceleration) at the endpoints, and (2) avoiding trap depth oscillations  at a frequency that could cause parametric heating\cite{Savard1997a}. 

To address (1), we  use the so-called ``minimum-jerk trajectory''~\cite{Shadmehr2005} to transport Cs. It is designed to translate the equilibrium point of a classical harmonic oscillator with minimal motional excitation. The displacement $x$ as a function of time t is given by 
$$x(t)=x_{\text{min jerk}}(t,d,T)=d\left(10(\frac{t}{T})^3-15(\frac{t}{T})^4+6(\frac{t}{T})^5\right)$$
where $d$ is the total distance traveled and $T$ is the total move time. 

However, the minimum jerk trajectory has a variable moving speed that is  problematic for constraint (2). Because the tweezer is transported by sweeping the RF frequency that drives the AOD in Fig.~\ref{app}, the trap depth oscillations arising from imperfections of the AOD (see Section~\ref{ssec:AODfringes}) would sweep through a band of frequencies and be more likely to excite a parametric heating resonance. 

Therefore, we devise a hybrid trajectory which uses constant velocity in the middle and minimum jerk at the endpoints. Thus, the oscillation frequency is constant for the middle part and the parameters can be more easily chosen to avoid parametric resonances. The displacement as a function of time for the hybrid trajectory is given by

\begin{widetext}
	\[
	x(t) = \left\{\begin{array}{lr}
	x_{\text{min jerk}}(t, 2 \Delta f,2\Delta t) & {\text{for } } 0\leq t\leq \Delta t\\
	\frac{15}{4}\frac{\Delta f}{2 \Delta t} & {\text{for } } \Delta t< t <T-\Delta t\\
	x_{\text{min jerk}}(t-T+ 2 \Delta t, 2 \Delta f,2\Delta t)+\alpha T \frac{15}{4}\frac{\Delta f}{2 \Delta t} & {\text{for } }T-\Delta t<t\leq T
	\end{array}\right\} 
	\]
\end{widetext}
where
$\Delta f= d/(2+\frac{15}{4}\frac{\alpha}{1-\alpha})$ and $\Delta t=\frac{1}{2}T(1-\alpha)$ are the distance covered and time elapsed, respectively, of the minimum jerk trajectory portion, and $\alpha$ is the fraction of the trajectory that is linear motion and can range from 0 (fully minimum jerk) to 1 (fully linear).

For data in Fig~\ref{merge}B, we use $d=2.5~\mu$m, $T=7.6$~ms, and $\alpha=0$. For the data in Fig~\ref{merge}C, we use $d=2.95~\mu$m and $\alpha=0.95$. 
We find the hybrid trajectory is more robust against parametric heating.

\section{Simulating merging of two tweezers}
To find the fastest speed at which we can merge single Na and
Cs atoms tweezers into the same tweezer, we simulate their time evolution using the split operator method~\cite{Tannor2007Book}.

The atomic polarizabilities are taken from Table 2 of Ref.~\cite{Safronova2006}.
The initial and final trap eigenfunctions are calculated with the Fourier Grid method~\cite{Kallush2006}. The ground state population at the end of the sequence is given by the squared overlap of the wavefunction following the time evolution with the ground state of the final trap.
The accuracy of these simulations is determined by the time step $\Delta t$ and position grid spacing $\Delta x$.  The accuracy of the split operator method is then set by $[T,V] \Delta t^2$, where $T$ and $V$ are the kinetic and potential energy operators, respectively. For the simulation data presented here, we use a time step of $\Delta t=0.1~\mu$s and spatial grid spacing $\Delta x$ = 1~nm, and have checked that the results of the simulation converge at these values.

The tweezer waist is estimated from scalar Gaussian beam propagation simulation of the input beam (whose waist we can measure), including the effect of the beam clipping on the objective aperture. The simulated electric field intensities along the radial and axial directions are fitted independently to those of a Gaussian beam. We find that doing so gives an input beam that is Gaussian except that the Rayleigh range is scaled by 1.39, to account for aberrations.

For the 976~nm tweezer, for 15~mW measured before a final beam expanding telescope, 9~mm waist input before the objective, the radial and axial waists at the tweezer are 0.844~$\mu$m and 4.875~$\mu$m ($z_R = 1.006~\mu$m) respectively. We match the calculated and measured radial and axial trapping frequencies of 125.7~kHz and 24.1~kHz respectively, by inserting a transmission coefficient T = 0.27 by hand. This includes transmission through many optical elements: dichroics, objective, glass cell, and electrode plate surfaces. Similarly, for the 700~nm tweezer, 6.6~mm input waist, 48~mW power before a final beam expanding  telescope, T= 0.36 gives 530.5~kHz and 92.7~kHz radial and axial trap frequencies, in good agreement with measurements.

\begin{figure}
	\includegraphics[width=\columnwidth]{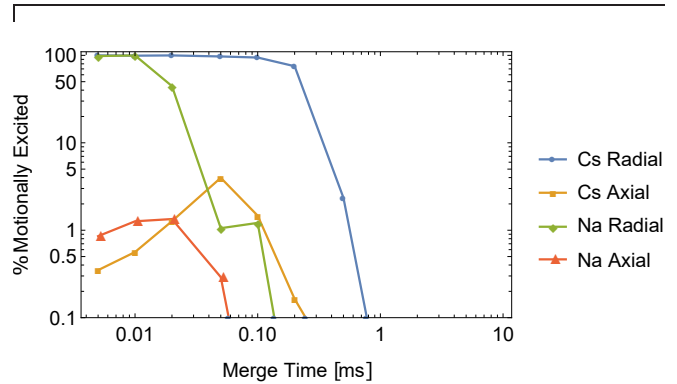}
	\caption{\textbf{Minimum merge time}. We numerically simulate the motional excitation as a function of merge time with fixed trap depth.}
	\label{SM10xMerge}
\end{figure}

By scanning the merge time and calculating the final wavefunction overlap with the motional ground state wavefunction, we find that we can scan more than $10\times$ faster (i.e. 2.95~$\mu$m in $<1$~ms) using a minimum jerk trajectory and still remain in the ground state with $>99.9\%$ probability (see Fig~\ref{SM10xMerge}), provided there are no technical imperfections.

\section{Derivation of trap depth oscillation frequency}
\label{ssec:AODfringes}

We use an IntraAction A2D-563AHF3.11 which can deflect the beam in two dimensions. The electro-optic medium is not angle-cut, and forms an acoustic cavity.  
The amplitude of the intracavity field affects the AOD diffraction efficiency and depends on RF drive frequency. Therefore, as the RF drive frequency is scanned to move the tweezer, the trap depth oscillates, in this case by 1\%. 

By scanning the tweezer position
along the merge axis and measuring the period of the intensity fringes, we measure the free spectral range of the acoustic cavity to be $FSR=97.5$~kHz. This is consistent with $FSR=v/2L$ where the length of the acousto-optic crystal $L\approx2$cm and the speed of sound is $v=3.63~mm/\mu$s. 

Scanning the RF drive frequency by 9.44~MHz moves the 976~nm trap 2.95~$\mu$m in the focal plane.

Therefore, the acoustic cavity causes the trap depth to oscillate at a frequency $v_{move} ~9.44$~MHz~$/(FSR\times 2.95~\mu$m), where $v_{move}$ is the speed at which the trap moves. For our hybrid trajectory in Section~\ref{joint_heat_section}, the trap depth oscillation during the linear part is therefore 9.9kHz.

\section{Simulated tweezer power 2D scan}

\begin{figure}
	\includegraphics[width=\columnwidth]{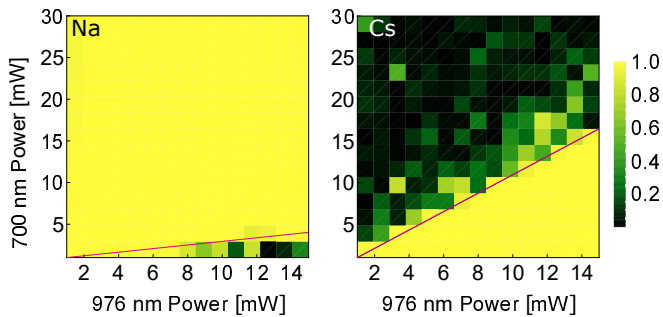}
	\caption{\textbf{Simulated 2D tweezer power scan}. Numerical simulation of the axial ground state population for Na and Cs following the merge sequence described in Sec.~\ref{joint_heat_section}. All the fundamental heating mechanisms (delineated by purple lines) are qualitatively reproduced.}
	\label{SM2DPowerScan}
\end{figure}

We perform a numerical simulation of the dynamics of merging atoms into one tweezer with different 700~nm and 976~nm tweezer powers. 
This yields the plots in Fig~\ref{SM2DPowerScan}A and B for Na and Cs, respectively. We find that heating regions arising from double-well for Na and anti-trapping of Cs are qualitatively reproduced (discrepancy in the exact size of the heating regions are attributed to aberrations of the tweezers which cause the actual trap depth to be different than expected).

We observe more overall heating in the experimental data compared to simulation, even in the regions that have no specific heating mechanism. This is likely caused by axial misalignment, which we estimate to be about 1.5~$\mu$m in this experiment. 

\section{Calibration of $\Omega_1$ and $\Omega_2$}
\label{app:calibrate_omega1_omega2}

We use the vector light shift of the 1038~nm beam on the Cs atom to calibrate the intensity. When the beam power $P_{PA}=20~$mW, the vector light shift was $\Delta_{VLS}=2\pi\times35.7$~kHz. We did PA to the $c^3\Sigma^+(v'=0)$ and measured a PA rate of $K_{PA}=1/0.35$~ms.

The excited state lifetime is assumed to be $\Gamma_e=1/30.4~$ns, the same as that of Cs $6^2P_{3/2}$.

Therefore $\Omega_1=\sqrt{\Gamma_e K_{PA}}=2\pi\times49~$kHz.

For arbitrary vector lightshift $\Delta_{VLS}$, $\Omega_1=\sqrt{\frac{\Delta_{VLS}}{2\pi\times35.7~kHz}}2\pi\times49~kHz$. 

Assuming no change in PA beam alignment, for arbitrary PA beam power $P_{PA}$, $\Delta_{VLS}$, $\Omega_1=\sqrt{\frac{P_{PA}}{20~mW}}2\pi\times49~kHz=2\pi\times11~kHz/\sqrt{mW}$. 

For the Raman transfer, we had a $\Delta_{VLS}=2\pi\times16.3~$kHz, so $\Omega_1=2\pi\times33~$kHz.

The theoretical FCF's yield $\Omega_2=292\times\Omega_1$.

\section{Using equal Raman beam powers}
\label{app:equal_beam_powers}
In the limits $d_2>>d_1$ (the bound-bound transition is much stronger than the free-bound transition) and $\Delta>>E_B/\hbar$, the differential AC stark shift and scattering rate are both proportional to $(P_1+P_2)d_2^2$.

\textit{1. Minimizing the number of photons scattered per $\pi-$time.}
The time required for a coherent transfer to occur is the so-called $\pi$-time $t_\pi$, where $\Omega_R t_\pi=\pi$. This is given by 

$$t_\pi=(2\pi\Delta)/(\sqrt{P_1 P_2}d_1 d_2)$$

where $d_i=\Omega_i/\sqrt{P_i}$ is proportional to the matrix element for the transition addressed by $\Omega_i$.

Therefore, we want to minimize $$(P_1+P_2)d_i^2/(\sqrt{P_1 P_2}d_1 d_2)$$
which occurs when $P_1=P_2$.

\textit{2. Minimizing the decoherence due to fluctuating differential AC Stark shift.} In the experiment, we assume that the total Raman beam power stability is some fixed fraction of the total power, $dP_{tot}\propto P_{tot}=P_1+P_2$. Differential fluctuations of the Raman resonance must be small compared to the spectral width of a coherent Raman transition, proportional to the Raman Rabi rate, $(\sqrt{P_1 P_2}d_1 d_2)/(2\Delta)$. Therefore, we also want to minimize $$(P_1+P_2)/(\sqrt{P_1 P_2})$$
As before, this occurs when $P_1=P_2$.

\section{Scattering rate of molecules due to tweezer}
\label{app:tweezer_v24_scattering_rate}
\begin{figure}
\centering
	\includegraphics[width=\columnwidth]{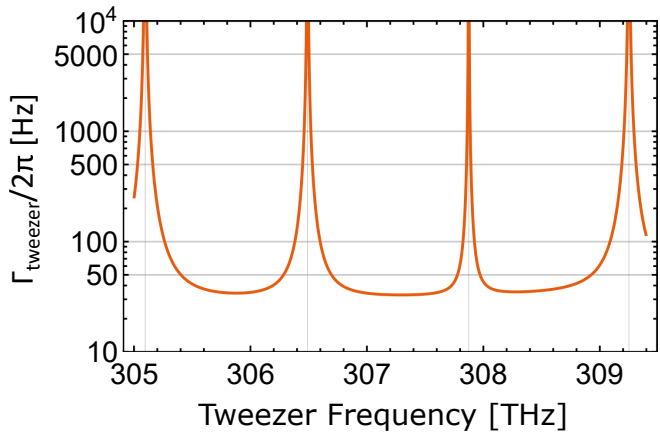}
	\caption{\textbf{Scattering rate due to the 15 mW  976~nm (307~THz) tweezer from $a^3\Sigma(v''=24)$}.   The calculation is performed using the $c^3\Sigma$ potential from Ref.~\cite{Grochola2011} and $a^3\Sigma$ potential from Ref.~\cite{Docenko2006}.  The peaks correspond to different vibrational states of the $c^3\Sigma$ potential.    } 
	\label{fig:tweezer_v24_scattering_rate}
\end{figure}

We calculate the scattering rate from the least-bound state  $v''= 24$ of the tweezer as a function of tweezer frequency in Fig.~\ref{fig:tweezer_v24_scattering_rate}.   The calculation assumes a tweezer beam power of 15~mW, and a beam waist of $0.8~\mu$m, and transmission through the objective and glass cell of 0.22.   The calculation includes vibrational level $v''= 24$ of the $a^3\Sigma^+$ ground state and a complete basis of vibrational eigenstates derived from the $c^3\Sigma^+$ excited state molecular potentials which are embedded in an isotropic harmonic well with a trap frequency of 80 kHz, which is the  geometric mean of the experiment axial and radial trapping frequencies.  The dipole matrix elements are assumed to be $3~e~a_0$ times the relevant wavefunction overlap.

\bibliography{master_ref}

\end{document}